\documentclass[aps,prl,preprint,superscriptaddress,showpacs]{revtex4-1}
\usepackage{amsmath}
\usepackage{graphicx}
\usepackage{mathrsfs}
\usepackage{dcolumn}
\usepackage{bm}
\usepackage{cases}
\usepackage{float}
\usepackage{multirow,booktabs}
\usepackage{makecell}
\usepackage{amsmath}
\usepackage{amsthm}
\usepackage{amssymb}
\usepackage{epstopdf}
\begin{document}


\title{Exploring sensitivity of charge-exchange ($p, n$) reactions to the neutron density distribution}

\author{Jian Liu}
\email{liujian@upc.edu.cn}
\affiliation{College of Science,  China University of Petroleum (East China), Qingdao 266580, China}
\affiliation{The Key Laboratory of High Precision Nuclear Spectroscopy, Institute of Modern Physics, Chinese Academy of Sciences}
\affiliation{Guangxi Key Laboratory of Nuclear Physics and Nuclear Technology, Guangxi Normal University}

\author{Yunsheng Wang}
\affiliation{College of Science, China University of Petroleum (East China), Qingdao 266580, China}
 
\author{Yonghao Gao}
\affiliation{School of Physics Science and Engineering, Tongji University, Shanghai 200092, China}

\author{Pawel Danielewicz}
\email{danielewicz@nscl.msu.edu}
\affiliation{Facility for Rare Isotope Beams and Department of Physics and Astronomy, Michigan State University, East Lansing, Michigan 48824, USA}

\author{Chang Xu}
\affiliation{Department of Physics, Nanjing University, Nanjing 210093, China}

\author{Zhongzhou Ren}
\affiliation{School of Physics Science and Engineering, Tongji University, Shanghai 200092, China}

\date{\today}

\begin{abstract}
	
\textbf{Background:} The determination of the nuclear neutron properties suffers from uncontrolled uncertainties, which attracted considerable attention recently, such as in the context of the PREX experiment.

\textbf{Purpose:}  Our aim is to analyze the sensitivity of charge-exchange ($p, n$) reactions to the neutron density distribution $\rho_{n}$ and constrain the neutron characteristics in the nuclear structure models.

\textbf{Method:}  By combing the folding and the mean-field models, the nucleon-nucleus ($NA$) potential can be obtained from the nuclear density distribution. Further, the ($p, p$) and ($p, n$) cross sections for $^{48}$Ca and $^{208}$Pb are calculated following the distorted-wave Born approximation (DWBA) method.

\textbf{Results:} Compared with the ($p, p$) cross section, the effects of $\rho_{n}$ variation on the ($p, n$) cross section are significant, which is due to the impact of isovector properties. Based on the global folding model analyses of data, it is found that $^{48}$Ca and $^{208}$Pb have relatively large neutron skin thickness $\Delta R_{n p}$.

\textbf{Conclusions:} Results illustrate that the charge-exchange ($p, n$) reaction is a sensitive probe of $\rho_{n}$. The results in this paper can offer useful guides for future experiments of neutron characteristics.

\end{abstract}


\maketitle

\section{\uppercase\expandafter{\romannumeral1}. Introduction}
The accurate description of neutron density distribution $\rho_{n}$ has been a longstanding problem in modern nuclear physics. Compared with the proton density distribution $\rho_{p}$, our knowledge of $\rho_{n}$ is very limited. The nuclear neutron characteristics are strongly connected with the equation of state (EOS) \cite{danielewicz2002determination,centelles2009nuclear}, the neutron star radius \cite{steiner2013neutron,tsang2020impact}, and the heavy ion collision \cite{tsang2012constraints,giacalone2020constraining}. In the last few years, different methods have been proposed and employed to probe $\rho_{n}$, such as the hadronic scattering \cite{patton2012neutrino,tagami2021neutron} and the formation of antiprotonic atoms \cite{trzcinska2001neutron,brown2007neutron,klos2007neutron}. However, the interpretation of these methods requires a model-dependent description of the strong interaction, leading to significant systematic besides statistical errors. It should be mentioned that the Lead Radius EXperiment (PREX) Collaboration at the Jefferson Laboratory (JLab) used the parity-violating electron scattering (PVES) to study $\rho_{n}$ for $^{208}$Pb  \cite{abrahamyan2012measurement,fattoyev2018neutron,adhikari2021accurate,reed2021implications,androic2022determination,adhikari2022new}. At present, $\rho_{n}$ is mainly measured through its contributions to the isoscalar properties. Compared with the isoscalar properties, the isovector properties better test uncertainties in $\rho_{n}$, therefore, it is extremely important to find an experimental observable of isovector properties. 

In the charge-exchange ($p, n$) reaction, the Fermi transitions $(\Delta L=0, \Delta S=0, \Delta T=1)$ between the initial state to isobaric analog states (IAS) provide a useful tool for studying isovector excitation. During the reaction process, the IAS essentially retains the same structure as the target nucleus, except for the replacement of a neutron by a proton \cite{zegers2003excitation,zegers2006t,loc2014charge,loc2017single}. The $NA$ potential can be written as the superposition of the isoscalar potential $U_{0}$ and isovector potential $U_{1}$ 
\begin{equation}\label{1}
	\begin{aligned}
      U(\bm{R})=U_{0}(\bm{R})+4 U_{1}(\bm{R}) \frac{\bm{t} \cdot \bm{T}}{A},
	\end{aligned}
\end{equation}
where \bm{$t$} and \bm{$T$} are the isospin of the projectile nucleon and the target nucleus, respectively. Compared with the $U_{0}$, the Lane potential $U_{1}$ is small, and its influence on the elastic scattering cross section is relatively limited \cite{satchler1983nuclear,huan2021excitation}. However, the $U_{1}$ reflects the differences between the neutron and proton potentials for elastic processes, and it determines the transition strength of the initial state to IAS in ($p, n$) reaction \cite{khoa2007folding}. Therefore, the charge-exchange ($p, n$) reactions can be a good probe of $\rho_{n}$. 

During the recent years, numerous models have been proposed to describe the isovector potential $U_{1}$. One such method is the optical model potential, which parameterizes the $U_{1}$ in Woods-Saxon form \cite{varner1991global,koning2003local}. However, the optical model parameters are derived from the elastic scattering data and do not connect to the nucleon-nucleon ($NN$) interaction \cite{satchler1979folding}. Efforts to describe $NN$ potential realistically at the microscopic level include the Argonne potential \cite{wiringa1995accurate,somasundaram2021constraints} and the Reid soft-core potential \cite{stoks1994construction}. Individual terms in a realistic $NN$ potential have a specific physical meaning but they do not directly relate to the nuclear density distribution or optical potential for scattering. For the purposes of relating the nucleon-nucleus scattering with the nuclear structure information, the folding model was developed in last decade \cite{khoa2002folding,khoa2007nuclear}. The folding model is built based on the effective $NN$ interaction \cite{deng2017improved,hamada2018single,durant2020dispersion}, which can be deduced from the \textit{G}-matrix elements of the Paris and Reid $NN$ potential, etc. \cite{khoa2016nuclear}. The folded potential is obtained by averaging the effective $NN$ interaction over the nuclear density distributions within the two colliding ions. If the effective $NN$ interaction is well defined, the folding model can provide a valid basis for study of $\rho_{n}$.

The neutron density distribution $\rho_{n}$ is usually calculated in a nuclear structure model, and there the self-consistent mean-field model for structure is a comprehensive and successful method to calculate the nuclear density distribution from the light to heavy nuclei \cite{kurasawa2019n,naito2021second,wang2021nucleon,niu2022effects}. Both relativistic and non-relativistic methods can be used to construct the mean-field model. For the binding energies $B/A$ and charge radii $R_{\mathrm{C}}$, the theoretical results of the mean-field model are consistent with the experimental data \cite{meucci2014neutron,liu2017coulomb,wang2020charge,wang2021global}. However, $\rho_{n}$ calculated from the mean-field models with different parameter sets vary considerably. The theoretical neutron skin thickness $\Delta R_{n p}$ given by the mean-field model range, in particular, from 0.1 fm to 0.32 fm for $^{208}$Pb \cite{roca2011neutron}. This is due to the lack of information on neutron characteristics when constraining the force parameters of mean-field model. Therefore, availability of suitable experimental observables of neutron characteristics is significant for the development of the nuclear structure model in general.

The main purpose of this paper is to analyze sensitivity of the charge-exchange ($p, n$) reactions to the neutron density distribution $\rho_{n}$. First, we study the nuclear properties of $^{208}$Pb and $^{48}$Ca in the Skyrme-Hartree-Fock (SHF) and the relativistic mean-field (RMF) frameworks. Next, we use the complex folding model and the hybrid folding model to generate $U_{0}$ and $U_{1}$ potentials in Eq. (\ref{1}), and further describe the ($p, p$) and ($p, n$) cross sections based on the distorted-wave Born approximation (DWBA) method \cite{frobrich1996theory}. Then, the renormalization coefficients of the folded potential are calibrated based on the experimental ($p, p$) and ($p, n$) cross sections of $^{208}$Pb and the $\Delta R_{n p}$ of PREX-II results. Finally, we explore the effects of $\rho_{n}$ on the ($p, n$) cross sections for the $^{208}$Pb. The calibrated renormalization coefficients are further substituted into calculations of ($p, p$) and ($p, n$) cross sections for $^{48}$Ca to investigate the neutron properties of $^{48}$Ca. The Calcium Radius EXperiment (CREX) plans to provide a measurement of the weak charge distribution and the neutron density of $^{48}$Ca \cite{horowitz2014electroweak}. The studies of quasielastic ($p, n$) reactions can offer useful guidance for the CREX experiment. Besides, the folding model analyses can also be used to study the $\alpha$ decay \cite{xu2008competition,deng2019significant}, the symmetry energy \cite{PhysRevC.81.064612,khoa2014folding} and the heavy ion collision \cite{bertulani2019introduction,danielewicz2022deblurring}.

This paper is organized as follows. In Sec.~II, the theoretical frameworks of the DWBA method, the folding model and the mean-field models are provided. In Sec.~III, the results and discussions of nuclear properties, and ($p, p$) and ($p, n$) cross sections for $^{208}$Pb and $^{48}$Ca are presented. Finally, conclusions are given in Sec.~IV.

\section{\uppercase\expandafter{\romannumeral2}. Theoretical framework}
In this section, we introduce the theoretical frameworks for calculating ($p, p$) and ($p, n$) scattering cross sections. First, we present the formulas for the ($p, p$) and ($p, n$) cross sections in the DWBA method. Then, we further investigate the $NA$ potential within the folding model. Finally, the corresponding formalisms for the SHF and RMF models are presented to calculate the density input for the folding model.

\subsection{A. DWBA cross sections} 
In the calculation of elastic scattering of charged particles, the cross section is obtained by considering both the Coulomb and nuclear scattering amplitudes. Correspondingly the ($p, p$) cross section can be decomposed into three terms \cite{bertulani2019introduction,danielewicz2017symmetry}
\begin{equation}\label{2}
	\begin{aligned}
		\frac{\mathrm{d} \sigma_{(p, p)}}{\mathrm{~d} \Omega}=\frac{\mathrm{d} \sigma_{C}}{\mathrm{~d} \Omega}+\frac{\mathrm{d} \sigma_{N}}{\mathrm{~d} \Omega}+\frac{\mathrm{d} \sigma_{i}}{\mathrm{~d} \Omega}
	\end{aligned}.
\end{equation}
Here, $d \sigma_{C} / d \Omega$ is the Rutherford cross section and  $d \sigma_{i} / d \Omega$ is the interference contribution. The remaining term is the nuclear cross section $d \sigma_{N} / d \Omega$, tied both to the Coulomb potential and the matrix element of the $NA$ potential in isospin space:
\begin{equation}\label{3}
	\begin{aligned}
		\langle \tau, Z|U(\boldsymbol{R})| \tau, Z\rangle=U_{0}(\boldsymbol{R}) \pm \frac{N-Z}{A} U_{1}(\boldsymbol{R}), \quad \text { with } \quad \tau=p,  n.
	\end{aligned}
\end{equation}
The $+$ sign of Eq. (\ref{3}) pertains to incident neutron and $-$ sign to incident proton. The angular structure of the nuclear cross section can be expressed as 
\begin{equation}\label{4}
	\begin{aligned}
		\frac{\mathrm{d} \sigma_{N}}{\mathrm{~d} \Omega}=\frac{1}{k^{2}} \frac{1}{2 s+1} \sum_{L}(2 L+1) \mathcal{A}_{L}^{N} P_{L}(\cos \theta),
	\end{aligned}
\end{equation}
where the expansion coefficients $\mathcal{A}_{L}^{N}$ are
\begin{equation}\label{5}
\begin{aligned}
	\mathcal{A}_{L}^{N} &=\frac{1}{4} \sum_{J^{\prime} \ell^{\prime}}\left(2 J^{\prime}+1\right)\left(2 \ell^{\prime}+1\right) \sum_{J \ell}(2 J+1)(2 \ell+1)\left(\begin{array}{lll}
		\ell & \ell^{\prime} & L \\
		0 & 0 & 0
	\end{array}\right)^{2} \\
	&\left.\times\left\{\begin{array}{ccc}
		\ell & \ell^{\prime} & L \\
		J^{\prime} & J & s
	\end{array}\right\}^{2} \operatorname{Re}\left[\mathrm{e}^{2 i\left(\sigma_{\ell}-\sigma_{\ell^{\prime}}\right.}\right)\left(S_{J^{\prime} \ell^{\prime}}^{N *}-1\right)\left(S_{J \ell}^{N}-1\right)\right].
\end{aligned}
\end{equation}
Here, $\sigma_{\ell}$ are Coulomb phase shifts and $S_{J \ell}^{N}$ are nuclear factors from solving Schr{\"o}dinger equation with the combination of Coulomb potential and nuclear potential in Eq. (\ref{3}). From Eq. (\ref{3}), it can be seen that the $U_{0}$ dominates the $NA$ potential, therefore, the ($p, p$) cross section mainly reflects the isoscalar properties of nucleus.

In ($p, n$) reaction, the matrix element that drives the transition from the initial state to the final state is
\begin{equation}\label{6}
	\begin{aligned}
		\langle n, Z+1|U(\boldsymbol{R})| p, Z\rangle=2 \frac{\sqrt{|N-Z|}}{A} U_{1}(\boldsymbol{R}).
	\end{aligned}
\end{equation}

In terms of Eq. (\ref{6}), the unpolarized ($p, n$) cross section in the DWBA approximation can be rewritten as \cite{frobrich1996theory,danielewicz2017symmetry}
\begin{equation}\label{7}
	\begin{aligned}
		\frac{\mathrm{d} \sigma_{(p, n)}}{\mathrm{d} \Omega}=(2 \pi)^{4} \mu_{p} \mu_{n} \frac{k_{n}}{k_{p}} \frac{1}{2 s+1} \sum_{M_{p} M_{n}}\left|2 \frac{\sqrt{|N-Z|}}{A} \int \mathrm{d} \boldsymbol{R} \chi_{n M_{n}}^{(-) \dagger}(\boldsymbol{R}) U_{1}(\boldsymbol{R}) \chi_{p M_{p}}^{(+)}(\boldsymbol{R})\right|^{2}.
	\end{aligned}
\end{equation}
Here $\mu$ and $k$ are reduced mass and center of mass (c.m.) wavevector in the $n$ or $p$ channels indicated with the subscript. The wave functions $\chi$ represent distorted waves of proton and neutron in the initial and final channels, which can be calculated in the consideration of the elastic scattering. The angular dependence of the ($p, n$) cross section can be expressed in a manner similar to Eq. (\ref{4})
\begin{equation}\label{8}
	\begin{aligned}
		\frac{\mathrm{d} \sigma_{(p, n)}}{\mathrm{d} \Omega}=\frac{1}{k_{p}^{2}} \frac{1}{2 s+1} \sum_{L}(2 L+1) \mathcal{A}_{L}^{(p, n)} P_{L}(\cos \theta).
	\end{aligned}
\end{equation}
Here, the coefficients $\mathcal{A}_{L}^{(p, n)}$ in the differential cross section are
\begin{equation}\label{9}
	\begin{aligned}
		\mathcal{A}_{L}^{(p, n)}=& 4 \mu_{p} \mu_{n} k_{p} k_{n} \sum_{J^{\prime} \ell^{\prime}}\left(2 J^{\prime}+1\right)\left(2 \ell^{\prime}+1\right) \sum_{J \ell}(2 J+1)(2 \ell+1) \\
		& \times\left(\begin{array}{lll}
			\ell & \ell^{\prime} & L \\
			0 & 0 & 0
		\end{array}\right)^{2}\left\{\begin{array}{ccc}
			\ell & \ell^{\prime} & L \\
			J^{\prime} & J & s
		\end{array}\right\}^{2} \operatorname{Re}\left[I_{J^{\prime} \ell^{\prime}}^{*} I_{J \ell}\right],
	\end{aligned}
\end{equation}
where $I$ are the partial-wave integrals
\begin{equation}\label{10}
	\begin{aligned}
I_{J \ell}=2 \frac{\sqrt{|N-Z|}}{A} \int_{0}^{\infty} \mathrm{d} R R^{2} u_{n J \ell}^{(+)}(R) U_{1}^{J \ell}(R) u_{p J \ell}^{(+)}(R),
	\end{aligned}
\end{equation}
and $u$ are radial wavefunctions for the initial and final channels.

\subsection{B. Folding model}\label{2.2}
The ($p, p$) and ($p, n$) scattering cross sections in Eqs. (\ref{2}) and (\ref{8}) are determined in terms of the $NA$ potential. In this paper, we use the folding model to calculate $U_{0}$ and $U_{1}$ and to connect the scattering cross sections and the nuclear structure model. In the folding model, the $NA$ potential $U_{\mathrm{N}}$ is evaluated as:
\begin{equation}\label{11}
	\begin{aligned}
		U_{\mathrm{N}}=\sum_{j \in A}\left[\left\langle i j\left|v_{D}\right| i j\right\rangle+\left\langle i j\left|v_{\mathrm{EX}}\right| j i\right\rangle\right],
	\end{aligned}
\end{equation}
where $v_{\mathrm{D}}$ and $v_{\mathrm{EX}}$ are the direct and exchange parts of the effective $NN$ interaction \cite{tan2021equation}. The spin-isospin term of the effective $NN$ interaction is decomposed as
\begin{equation}\label{12}
	\begin{aligned}
		\begin{aligned}
			v_{D(\mathrm{EX})}(\rho, E, s)=& v_{00}^{D(\mathrm{EX})}(\rho, E, s)+v_{10}^{D(\mathrm{EX})}(\rho, E, s)\left(\boldsymbol{\sigma} \cdot \boldsymbol{\sigma}^{\prime}\right) \\
			&+v_{01}^{D(\mathrm{EX})}(\rho, E, s)\left(\boldsymbol{\tau} \cdot \boldsymbol{\tau}^{\prime}\right)+v_{11}^{D(\mathrm{EX})}(\rho, E, s)\left(\boldsymbol{\sigma} \cdot \boldsymbol{\sigma}^{\prime}\right)\left(\boldsymbol{\tau} \cdot \boldsymbol{\tau}^{\prime}\right).
		\end{aligned}
	\end{aligned}
\end{equation}
Here, $s$ is the distance between a target nucleon and the incident proton, and $\rho$ is the nuclear density. The contribution from the spin dependent terms ($v_{10}$ and $v_{11}$) in Eq. (\ref{12}) is exactly zero for a spin-saturated target. In using the explicit $\rho_{p}$ and $\rho_{n}$ as the input of folding model, the HF potential $U_{\mathrm{N}}$ can be separated into the isoscalar ($U_{\mathrm{IS}}$) and isovector ($U_{\mathrm{IV}}$) parts as 
\begin{equation}\label{13}
	\begin{aligned}
		U_{\mathrm{N}}(E, \boldsymbol{R})=U_{\mathrm{IS}}(E, \boldsymbol{R}) \pm U_{\mathrm{IV}}(E, \boldsymbol{R}),
	\end{aligned}
\end{equation}
where the ($+$) and ($-$) refer to neutrons and protons, respectively \cite{loan2020rearrangement}. For the complex effective $NN$ interaction, the $U_{\mathrm{IS(IV)}}$ should be calculated explicitly in terms of real ($V_{\mathrm{IS(IV)}}$) and imaginary ($W_{\mathrm{IS(IV)}}$) parts as \cite{khoa2014folding}
\begin{equation}\label{14}
	\begin{aligned}
		U_{\mathrm{IS}(\mathrm{IV})}(E, \boldsymbol{R})=V_{\mathrm{IS}(\mathrm{IV})}(E, \boldsymbol{R})+i W_{\mathrm{IS}(\mathrm{IV})}(E, \boldsymbol{R}).
	\end{aligned}
\end{equation}
In the spirit of Eq. (\ref{11}), the individual terms in Eq. (\ref{14}) may be calculated from
\begin{equation}\label{15}
	\begin{aligned}
\begin{aligned}
	V^{\mathrm{IS(IV)}}(E, \boldsymbol{R}) &= \int F_{\mathrm{IS(IV)}}^{V}(E, \rho)\left\{\left[\rho_{\mathrm{n}}(\boldsymbol{r}) \pm \rho_{\mathrm{p}}(\boldsymbol{r})\right] v_{\mathrm{IS(IV)}}^{\mathrm{D}}(s)\right.\\
	&\left.+\left[\rho_{n}(\boldsymbol{R}, \boldsymbol{r}) \pm \rho_{p}(\boldsymbol{R}, \boldsymbol{r})\right] v_{\mathrm{IS(IV)}}^{\mathrm{EX}}(s) j_{0}(k(E, R) s)\right\} d^{3} r,
\end{aligned}
	\end{aligned}
\end{equation}
\begin{equation}\label{16}
	\begin{aligned}
	\begin{aligned}
		W^{\mathrm{IS(IV)}}(E, \boldsymbol{R}) &= \int F_{\mathrm{IS(IV)}}^{W}(E, \rho)\left\{\left[\rho_{n}(\boldsymbol{r}) \pm \rho_{p}(\boldsymbol{r})\right] v_{\mathrm{IS(IV)}}^{\mathrm{D}}(s)\right.\\
		&\left.+\left[\rho_{n}(\boldsymbol{R}, \boldsymbol{r}) \pm \rho_{p}(\boldsymbol{R}, \boldsymbol{r})\right] v_{\mathrm{IS(IV)}}^{\mathrm{EX}}(s) j_{0}(k(E, R) s)\right\} d^{3} r.
	\end{aligned}
	\end{aligned}
\end{equation}
Here, the ($+$) refer to isoscalar and ($-$) to isovector, and $\boldsymbol{s}=\boldsymbol{R}-\boldsymbol{r}$ is the folding distance. The functions $v_{\mathrm{IS(IV)}}^{\mathrm{D(EX)}}(s)$ represent the radial shapes of the isoscalar and isovector $NN$ interactions, that get deduced from the \textit{G}-matrix elements of the realistic $NN$ potential \cite{anantaraman1983effective}. The factors $F_{\mathrm{IS(IV)}}^{u}(E, \rho)$ represent the density dependence for the real part ($u=V$) and imaginary part ($u=W$) of the potentials, spelled out later in this paper. The local momentum of relative motion $k(E, R)$ is determined from:
\begin{equation}\label{17}
	\begin{aligned}
	k^{2}(E, R)=\frac{2 \mu}{\hbar^{2}}\left[E_{\mathrm{c.m.}}-U_{\mathrm{N}}(E, R)-U_{\mathrm{C}}(R)\right].
	\end{aligned}
\end{equation}
Here, $U_{\mathrm{C}}(R)$ and $U_{\mathrm{N}}(E, R)$ are the Coulomb potential and the real $NA$ potential, respectively. In this paper, the exchange parts of both the $U_{\mathrm{IS}}$ and $U_{\mathrm{IV}}$ are evaluated iteratively using the finite-range exchange interaction, which is more accurate than those given by a zero-range approximation for the exchange term. Combining the Eqs. (\ref{15})-(\ref{17}), we can get the self-consistent $U_{\mathrm{N}}$ by the iterative solution finally.

\subsection{C. Nuclear density distribution}
The self-consistent mean-field model is a microscopic and successful model frequently employed in the context of nuclear structure. There are two dominant approaches to the mean-field: the nonrelativistic and relativistic. In the following, we introduce the theoretical frameworks for the nonrelativistic Skyrme-Hartree-Fock (SHF) and the relativistic mean-field (RMF) models.

\subsubsection{{\romannumeral1}. Nonrelativistic Skyrme-Hartree-Fock method }
Within the SHF method, the energy density functional $H(\mathbf{r})$ can be written as \cite{stoitsov2007variation,wang2020charge}
\begin{equation}\label{18}
	\begin{aligned}
		\begin{aligned}
			H(\mathbf{r})=& \frac{\hbar^{2}}{2 m} \tau+\frac{1}{2} t_{0}\left[\left(1+\frac{1}{2} x_{0}\right) \rho^{2}-\left(\frac{1}{2}+x_{0}\right) \sum_{q} \rho_{q}^{2}\right] \\
			&+\frac{1}{2} t_{1}\left[\left(1+\frac{1}{2} x_{1}\right) \rho\left(\tau-\frac{3}{4} \Delta \rho\right)-\left(\frac{1}{2}+x_{1}\right) \sum_{q} \rho_{q}\left(\tau_{q}-\frac{3}{4} \Delta \rho_{q}\right)\right] \\
			&+\frac{1}{2} t_{2}\left[\left(1+\frac{1}{2} x_{2}\right) \rho\left(\tau+\frac{1}{4} \Delta \rho\right)-\left(\frac{1}{2}+x_{2}\right) \sum_{q} \rho_{q}\left(\tau_{q}+\frac{1}{4} \Delta \rho_{q}\right)\right] \\
			&+\frac{1}{12} t_{3} \rho^{\alpha}\left[\left(1+\frac{1}{2} x_{3}\right) \rho^{2}-\left(x_{3}+\frac{1}{2}\right) \sum_{q} \rho_{q}^{2}\right] \\
			&-\frac{1}{8}\left(t_{1} x_{1}+t_{2} x_{2}\right) \sum_{i j} \mathbf{J}_{i j}^{2}+\frac{1}{8}\left(t_{1}-t_{2}\right) \sum_{q, i j} \mathbf{J}_{q, i j}^{2}-\frac{1}{2} W_{0} \sum_{i j k} \varepsilon_{i j k}\left[\rho \nabla_{k} \mathbf{J}_{i j}+\sum_{q} \rho_{q} \nabla_{k} \mathbf{J}_{q, i j}\right],
		\end{aligned}
	\end{aligned}
\end{equation}
where the $\rho(\mathbf{r})$, $\tau(\mathbf{r})$ and $\mathbf{J}_{i j}(\mathbf{r})$ represent the local partical density, kinetic energy density and spin-orbit density, and the different parameters are adjusted to yield desired nuclear properties. The index $q$ refers to neutrons and protons.

The Hartree-Fock (HF) equation is derived from the variation of total energy with respect to single-particle orbitals $\Phi_{\alpha}^{q}(\boldsymbol{r})$. By iteratively solving the HF equation, the nuclear density distributions can be obtained:
\begin{equation}\label{19}
	\begin{aligned}
\rho_{q}(\boldsymbol{r})=\sum_{\alpha}|\Phi_{\alpha}^{q}(\boldsymbol{r})|^{2}.
	\end{aligned}
\end{equation}

\subsubsection{{\romannumeral2}. Relativistic mean-field method }
In the framework of RMF method \cite{todd2005neutron,liu2017coulombb}, the starting point is the Lagrangian density:
\begin{equation}\label{20}
	\begin{aligned}
		\begin{aligned}
			\mathcal{L}=& \bar{\Psi}\left(i \gamma^{\mu} \partial_{\mu}-M\right) \Psi-g_{\sigma} \bar{\Psi} \sigma \Psi-g_{\omega} \bar{\Psi} \gamma^{\mu} \omega_{\mu} \Psi-g_{\rho} \bar{\Psi} \gamma^{\mu} \rho_{\mu}^{a} \tau^{a} \Psi+\frac{1}{2} \partial^{\mu} \sigma \partial_{\mu} \sigma \\
			&-\frac{1}{2} m_{\sigma}^{2} \sigma^{2}-\frac{1}{3} g_{2} \sigma^{3}-\frac{1}{4} g_{3} \sigma^{4}-\frac{1}{4} \Omega^{\mu \nu} \Omega_{\mu \nu}+\frac{1}{2} m_{\omega}^{2} \omega^{\mu} \omega_{\mu}+\frac{1}{4} c_{3}\left(\omega_{\mu} \omega^{\mu}\right)^{2} \\
			&-\frac{1}{4} \vec{R}^{\mu \nu} \cdot \vec{R}_{\mu \nu}+\frac{1}{2} m_{\rho}^{2} \bar{\rho}^{\mu} \cdot \vec{\rho}_{\mu}-\frac{1}{4} F^{\mu \nu} F_{\mu \nu}-e \bar{\Psi} \gamma^{\mu} A_{\mu} \frac{1}{2}\left(1-\tau^{3}\right) \Psi,
		\end{aligned}
	\end{aligned}
\end{equation}
where $\sigma$, $\omega$ and $\rho$ represent the isoscalar-scalar, isoscalar-vector and isovector-vector mesons, respectively.

Under the no-sea approximation and mean-field approximation, the Dirac equation for nucleons and the Klein-Gordon equations for meson fields can be obtained from the variational principle. By solving the motion equation iteratively, we can obtain the large component $f$ and small component $g$ of the nucleon wave function $\psi$ and derive the nucleon density:
\begin{equation}\label{21}
	\begin{aligned}
		\rho_{q}(\boldsymbol{r})=\sum_{\alpha}\left(\left|f_{\alpha}^{q}(\boldsymbol{r})\right|^{2}+\left|g_{\alpha}^{q}(\boldsymbol{r})\right|^{2}\right).
	\end{aligned}
\end{equation}
The SHF and RMF codes used in this paper allow for axially symmetry deformations \cite{stoitsov2007variation,todd2005neutron}, although these are not important in the present work.

\section{\uppercase\expandafter{\romannumeral3}. Numerical results and discussions}\label{sec:III}
In this section, we focus on the sensitivities of ($p, p$) and ($p, n$) scattering cross sections to the neutron density distribution $\rho_{n}$. We first investigate the binding energies per nucleon $B/A$, charge root-mean-square (RMS) radii $R_{\mathrm{C}}$ and neutron skin thickness $\Delta R_{n p}$ for different interactions. Next, we calculate the ($p, p$) and ($p, n$) cross sections at 35$\,$MeV and 45$\,$MeV within the complex folding and hybrid folding models. The $^{48}$Ca and $^{208}$Pb nuclei are chosen to illustrate our points.

\subsection{A. Ground-state properties of $^{208}$Pb and $^{48}$Ca}
\label{Sec:A}
In this subsection, the binding energies per nucleon $B/A$, charge RMS radius $R_{\mathrm{C}}$ and neutron skin $\Delta R_{n p}$ calculated in the RMF and SHF models with different interaction parameter are presented. Recently, the PREX-I and the PREX-II results for $^{208}$Pb have been reported in Refs. \cite{abrahamyan2012measurement,adhikari2021accurate}, including skin values of $\Delta R_{n p}^{\text{PREX-I}}=0.33_{-0.18}^{+0.16} \, \mathrm{fm}$ and $\Delta R_{n p}^{\text{PREX-II}}=0.283_{-0.071}^{+0.071} \, \mathrm{fm}$, respectively. For investigations in this work, we choose the NL3$^*$, NL1, SkO and SLy4 parameter sets in the RMF and SHF models for calculating the nuclear ground-state properties. The $\Delta R_{n p}$ results of NL3$^*$, NL1 and SkO correspond to the central value, upper and lower limit of the PREX-II skin, respectively, and the $\Delta R_{n p}$ result of SLy4 corresponds to the lower limit of the PREX-I skin. Our aforementioned theoretical results are represented in Table I. As might be expected, $B/A$ and $R_{\mathrm{C}}$ of $^{48}$Ca and $^{208}$Pb calculated with different parameter sets agree well with data, such as at the level of $0.5 \%$ for $^{208}$Pb. This is because the isoscalar predictions of the mean-field models have been historically well constrained with the existing experimental data. 

\begin{table}[htbp]
	\begin{center}
		\caption{Binding energies per nucleon $B/A$, charge RMS radii $R_{\mathrm{C}}$ and neutron skin $\Delta R_{n p}$ calculated with different parameter sets of the SHF and RMF models. Experimental data are from Refs. \cite{angeli2013table,adhikari2021accurate,wang2021ame}.}
		\label{Table:I}
		\begin{tabular}{llcllcllcllc}
			\hline
			\hline
			\multicolumn{1}{c}{Nucleus} &  & Parameter &  &  & $B/A$ (MeV)   &  &  & $R_{\mathrm{C}}$ (fm)  &  &  & $\Delta R_{n p}$ (fm)  \\ \hline
			$^{48}$Ca                        &  & SLy4           &  &  & 8.71 &  &  & 3.544  &  &  & 0.153 \\
			&  & SkO           &  &  & 8.51 &  &  & 3.511  &  &  & 0.248 \\
			&  & NL3$^*$           &  &  & 8.62 &  &  & 3.527  &  &  & 0.246 \\
			&  & NL1            &  &  & 8.60 &  &  & 3.549  &  &  & 0.271 \\
			&  & Expt.          &  &  & 8.67 &  &  & 3.477  &  &  &       \\
			$^{208}$Pb                       &  & SLy4           &  &  & 7.86 &  &  & 5.517  &  &  & 0.160 \\
			&  & SkO           &  &  & 7.83 &  &  & 5.510  &  &  & 0.218 \\
			&  & NL3$^*$           &  &  & 7.88 &  &  & 5.518  &  &  & 0.284 \\
			&  & NL1            &  &  & 7.89 &  &  & 5.537  &  &  & 0.313 \\
			&  & Expt.          &  &  & 7.87 &  &  & 5.501  &  &  & $0.283 \pm 0.071$ \\ \hline\hline
		\end{tabular}
	\end{center}
\end{table}

Contrary to the binding energies per nucleon $B/A$ and charge RMS radii $R_{\mathrm{C}}$, there are large variations in the $\Delta R_{n p}$ between different nuclear structure models and parameter sets. This can be attributed to variations in the isovector interaction, which is poorly constrained due to the historical lack of sufficiently precise experimental data on neutron properties. Although PREX-II has reported the updated neutron radius $R_{n}$ for $^{208}$Pb with a precision of virtually $1.0 \%$, its error bar covers the theoretical results of many mean-field parameter sets. In Fig. 1, we present the ground-state $\rho_{n}$ and $\rho_{p}$ of $^{208}$Pb generated with different parameter sets. Variations in theoretical $\rho_{n}$ corresponding to the error bar of PREX-II result are shown in the shaded part in this figure. One can observe that variations in $\rho_{p}$ are generally more modest and especially in the outer region that gets weighted by $r^{2}$ factor in calculations of any expectation values. By contrast, $\rho_{n}$ has a large variation in the outer region under the error bar of the PREX-II result.

\begin{figure}[htbp]
	\centering
	\includegraphics[width=8.5cm,angle=0,clip=true]{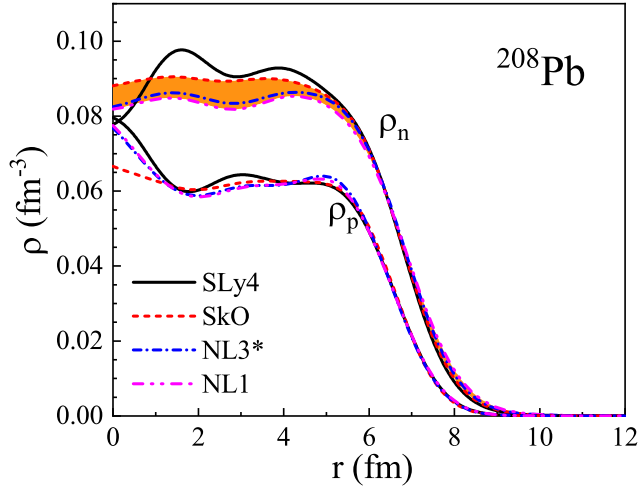}
	\caption{(Color online) Ground-state $\rho_{n}$ and $\rho_{p}$ for $^{208}$Pb calculated by the various models. The shaded part is shown to reproduce the experimental error bar of PREX-II data.}
	\label{fig:1}
\end{figure}

Besides the PVES experiment, the quasielastic ($p, n$) scattering is also sensitive to the nuclear isovector properties. Therefore, that scattering can be used to test $\Delta R_{n p}$ \cite{loc2014charge,loc2017single}. Form Eqs. (\ref{15}) and (\ref{16}), one can see that $U_{\mathrm{IV}}$ is directly related to the $\Delta R_{n p}$. However, the renormalization coefficients of the folded potential are undetermined in the calculation of scattering cross section. In the next part, we constrain the renormalization coefficients based on the $\Delta R_{n p}$ of PREX-II results and the experimental data of the ($p, p$) and quasielastic ($p, n$) cross sections on  $^{208}$Pb. With the fine-tuned folded potential, we further study the sensitivities of the ($p, p$) and ($p, n$) cross sections to the neutron density distribution $\rho_{n}$.

\subsection{B. Complex folding model analysis}

Next, we examine the ($p, p$) and ($p, n$) scattering cross sections within the complex folding model. The basic inputs for the folding model are the nuclear density distribution and the effective $NN$ interaction. The nuclear densities for Eqs. (\ref{15}) and (\ref{16}) are obtained from the mean-field models. For the effective $NN$ interaction, we choose the CDM3Y6 interaction  \cite{khoa2007folding}. The real part of the isoscalar density dependence of CDM3Y6 interaction $F_{\mathrm{IS}}^{\scriptscriptstyle V}(E, \rho)$ can be expressed as
\begin{equation}\label{22}
	\begin{aligned}
		F_{\mathrm{IS}}^{\scriptscriptstyle V}(E, \rho)=g(E) C_{0}\left[1+\alpha_{0} \exp \left(-\beta_{0} \rho\right)-\gamma_{0} \rho\right],
	\end{aligned}
\end{equation}
where the parameters combination $C_{0}$, $\alpha_{0}$, $\beta_{0}$ and $\gamma_{0}$ provides a nuclear incompressibility of $K \approx 252$ MeV \cite{khoa2002folding}. The energy dependence of $F_{\mathrm{IS}}^{\scriptscriptstyle V}(E, \rho)$ is contained in the factor $g$ changing linearly with energy $g(E)$$\approx1-0.002E$. Given the successful application of such parametrized density dependence in numerous folding calculations, the imaginary part of such isoscalar density dependence $F_{\mathrm{IS}}^{\scriptscriptstyle W}(E, \rho)$ and isovector density dependence $F_{\mathrm{IV}}^{u}(E, \rho)$ are assumed to have the form inspired by $F_{\mathrm{IS}}^{\scriptscriptstyle V}(E, \rho)$
\begin{equation}\label{23}
	\begin{aligned}
		F_{\mathrm{IS}}^{\scriptscriptstyle W}(E, \rho)=C_{0}^{\scriptscriptstyle W}(E)\left[1+\alpha_{0}^{\scriptscriptstyle W}(E) \exp \left(-\beta_{0}^{\scriptscriptstyle W}(E) \rho\right)-\gamma_{0}^{\scriptscriptstyle W}(E) \rho\right],
	\end{aligned}
\end{equation}
\begin{equation}\label{24}
	\begin{aligned}
		F_{\mathrm{IV}}^{u}(E, \rho)=C_{1}^{u}(E)\left[1+\alpha_{1}^{u}(E) \exp \left(-\beta_{1}^{u}(E) \rho\right)-\gamma_{1}^{u}(E) \rho\right],
	\end{aligned}
\end{equation}
in which the parameters of $F_{\mathrm{IS}}^{\scriptscriptstyle W}(E, \rho)$ and $F_{\mathrm{IV}}^{u}(E, \rho)$ are assumed to be energy-dependent and are adjusted at each incident energy $E$. In Eqs. (\ref{15}) and (\ref{16}), the radial shapes of direct and exchange parts $v^{\mathrm{D(EX)}}$ of CDM3Y6 interaction are taken from the M3Y-Paris interaction as a combination of three Yukawa terms \cite{anantaraman1983effective}
\begin{equation}\label{25}
	\begin{aligned}
	v_{\mathrm{IS(IV)}}^{\mathrm{D}(\mathrm{EX})}(s)=\sum_{v=1}^{3} Y_{\mathrm{IS(IV)}}^{\mathrm{D}(\mathrm{EX})}(v) \frac{\exp \left(-R_{v} s\right)}{R_{v} s},
	\end{aligned}
\end{equation}
where the Yukawa strengths can be found in Ref. \cite{khoa2007folding}.

With Eqs. (\ref{22})-(\ref{25}), the $V_{\mathrm{IS(IV)}}$ and $W_{\mathrm{IS(IV)}}$ of the folded potential in Eqs. (\ref{15}) and (\ref{16}) can be calculated explicitly and the $NA$ potential can be evaluated as
\begin{equation}\label{26}
	\begin{aligned}
		U_{\mathrm{N}}(R)=N_{V}\left[V_{\mathrm{IS}}(R) \pm N_{V 1} V_{\mathrm{IV}}(R)\right]+i N_{W}\left[W_{\mathrm{IS}}(R) \pm W_{\mathrm{IV}}(R)\right],
	\end{aligned}
\end{equation}
where the ($+$) and ($-$) refer to neutrons and protons, respectively. The $N_{V(W)}$ and $N_{V1}$ are the renormalization coefficients established in this paper. The $N_{V(W)}$ and $N_{V1}$ are calibrated based on the experimental data for the ($p, p$) and ($p, n$) cross sections, assuming validity of the central $\Delta R_{n p}$ value from PREX-II. The $N_{V1}$ is further tuned for different nuclei. The transition matrix element of Eq. (\ref{6}) can be further expressed in terms of the folded potential $U_{\mathrm{IV}}$ as \cite{khoa2014folding}

\begin{equation}\label{27}
	\begin{aligned}   
     	\langle n, Z+1|U(R)| p, Z\rangle &=2\frac{\sqrt{|N-Z|}}{A} U_1(R) \\
     	&=\frac{2}{\sqrt{|N-Z|}}U_{\mathrm{IV}}(R) .
	\end{aligned}
\end{equation}
During the calibration process, $\rho_{n}$ is calculated using the NL3$^*$ parameter set, because it gives a $\Delta R_{n p}$ consistent with the central value of the PREX-II $\Delta R_{n p}$ results. The best-fit renormalization coefficients at the incident energies of 35$\,$MeV and 45$\,$MeV are listed in Table \ref{Table:III}. The corresponding parameters of CDM3Y6 interaction for incident energies at 35$\,$MeV and 45$\,$MeV are taken from Refs. \cite{khoa2007folding,communication}. Finally, the net scattering potential is obtained from the superposition of the $NA$ potential $U_{\mathrm{N}}$, the spin-orbital potential $U_{\mathrm{LS}}$ and the Coulomb potential $U_{\mathrm{C}}$.

\begin{table}[htbp]
	\begin{center}
		\caption{Renormalization coefficients $N_{V(W)}$ and $N_{V1}$ of the complex folded potential Eq. (\ref{26}) at 35$\,$MeV and 45$\,$MeV, which are calibrated based on the experimental ($p, p$) and ($p, n$) cross sections, assuming that the central $\Delta R_{n p}$ value from PREX-II is valid.}
		\label{Table:III}
\begin{tabular}{ccccclcllcllc}
	\hline\hline
	$E$  &  &  & $N_{V}$    &  &  & $N_{W}$    &  &  & $N_{V1}$($^{48}$Ca)   &  &  & $N_{V1}$($^{208}$Pb)   \\ \hline
	35 &  &  & 0.849 &  &  & 0.591 &  &  & 0.992 &  &  & 1.749 \\
	45 &  &  & 0.840 &  &  & 0.619 &  &  & 1.136 &  &  & 1.452 \\ \hline\hline
\end{tabular}
	\end{center}
\end{table}

\begin{figure}[htbp]
\centering
  \includegraphics[width=8.5cm,angle=0,clip=true]{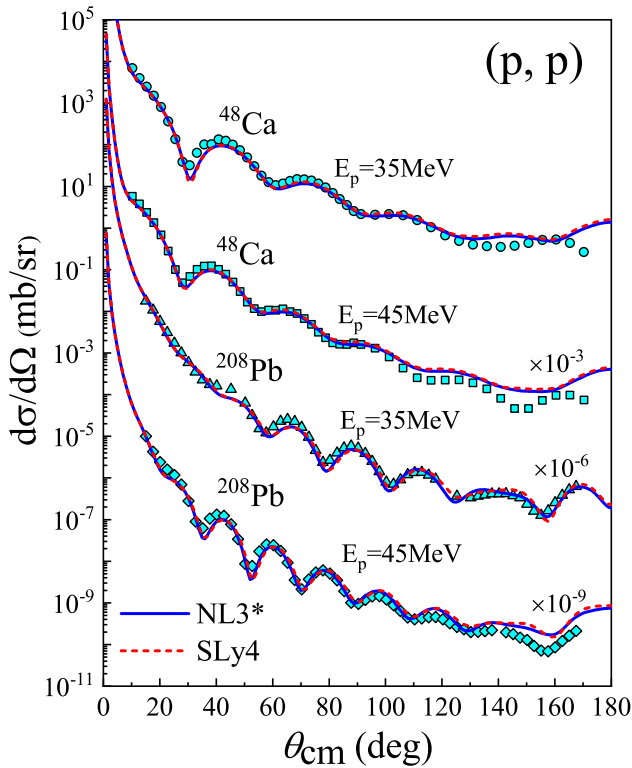}\\
  \caption{(Color online) Different ($p, p$) cross sections on $^{48}$Ca and $^{208}$Pb targets at 35$\,$MeV and 45$\,$MeV from the calculations with the complex folded potential of Eq. (\ref{26}), based on the nuclear densities calculated by the SHF and RMF models. The experimental data stem from Refs. \cite{mccamis1986elastic,van1974optical}.}
  \label{fig:2}
\end{figure}

The different ($p, p$) cross sections on $^{208}$Pb calculated with the complex folded potential of Eq. (\ref{26}), at 35$\,$MeV and 45$\,$MeV, are shown in Fig. \ref{fig:2}. It can be seen that the complex folded potential gives good ($p, p$) descriptions on cross section, which confirms the reliability of the complex folding model, especially here of its isoscalar component $U_{\mathrm{IS}}$. To provide insights, the ($p, p$) cross sections are obtained using both nuclear density distributions calculated with the NL3$^*$ and SLy4 interaction. Note that $\Delta R_{n p}$ calculated with these two interactions is different in Table \ref{Table:I}, but the difference is hardly reflected in the ($p, p$) cross sections. This is because the ($p, p$) cross section is primarily related to the isoscalar net density, and only weakly to isovector density. 

We further present the ($p, p$) cross sections on $^{48}$Ca in Fig. \ref{fig:2}, again using renormalization coefficients from Table \ref{Table:III}. One can see that the theoretical results are in a reasonable agreement with experimental data. Importantly, the isoscalar renormalization coefficients used for $^{208}$Pb are reliable in calculating the ($p, p$) cross sections for the other nucleus. Similarly to $^{208}$Pb, little difference is observed when in the ($p, p$) cross sections of $^{48}$Ca are calculated for different $\Delta R_{n p}$. Concluding, while the ($p, p$) scattering can test the net density of the nucleus, it is not very sensitive to $\Delta R_{n p}$.

\begin{figure}[htbp]
	\centering
	\includegraphics[width=8.5cm,angle=0,clip=true]{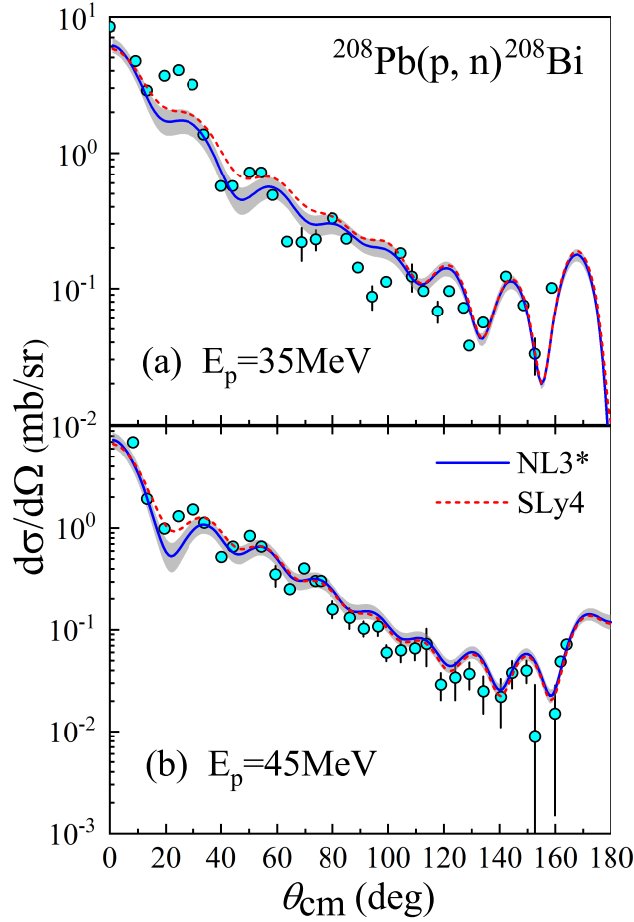}\\
	\caption{(Color online) Cross section for the quasielastic $^{208}$Pb($p, n$)$^{208}$Bi reaction at 35 (a) and 45 (b) MeV. The experimental data from Ref. \cite{doering1975microscopic} are represented by circles. DWBA calculations in the folding model of Eq. (\ref{26}) are represented by lines, solid for the NL3$^*$ interaction and dashed for SLy4. The shaded region represents the span of NL3$^*$ results when the neutron radii corresponding to the nominal uncertainty in PREX-II.}
	\label{fig:3}
\end{figure}

The ($p, n$) cross sections on $^{208}$Pb obtained using NL3$^*$ and SLy4 interactions at 35$\,$MeV and 45$\,$MeV are presented in Fig. \ref{fig:3}. It can be seen that the calculations reproduce the general trend of the ($p, n$) experimental data, which demonstrates general validity of the isovector part $U_{\mathrm{IV}}$ of the $NA$ potential. There are evident differences between the predictions from these two models in the region $\theta=20^{\circ} \mbox{-} 80^{\circ}$, which indicates that the effects of isovector density on ($p, n$) reaction are more obvious than on ($p, p$). This is because the ($p, n$) cross section is dominated by the $U_{\mathrm{IV}}$ component, which is connected to the nuclear isovector density and, thus, magnifies the effects of $\Delta R_{n p}$. 

As the $\Delta R_{n p}$ calculated for NL3$^*$ corresponds just to the central value of the $\Delta R_{n p}$ result of PREX-II, we can explore the whole range of PREX-II uncertainty by stretching $\rho_{n}$ from NL3$^*$ with a factor $\lambda$  \cite{liu2013theoretical}, i.e., carrying out transformation for the neutron density $\rho_{n}(r) \rightarrow \lambda^{-3} \rho_{n}\left(r / \lambda\right)$. With this method, the neutron radius is scaled by $\lambda$:
\begin{equation}
	\begin{aligned}
		R_{n}^{\prime}=\sqrt{\int 4 \pi r^{4} \frac{1}{\lambda^{3}} \rho_{n}\left(\frac{r}{\lambda}\right) d r}=\lambda \cdot R_{n}.\nonumber
	\end{aligned}
\end{equation}
By choosing different $\lambda$, we can span the full range of nominal uncertainty for the PREX-II $R_{n}$ result, and the corresponding ($p, n$) cross sections are shown by the shaded areas in Fig. \ref{fig:3}. One can see that the effects on ($p, n$) caused by the modifications of $\rho_{n}$ are significant over the uncertainty of PREX-II result.

\begin{figure}[htbp]
	\centering
	\includegraphics[width=8.5cm,angle=0,clip=true]{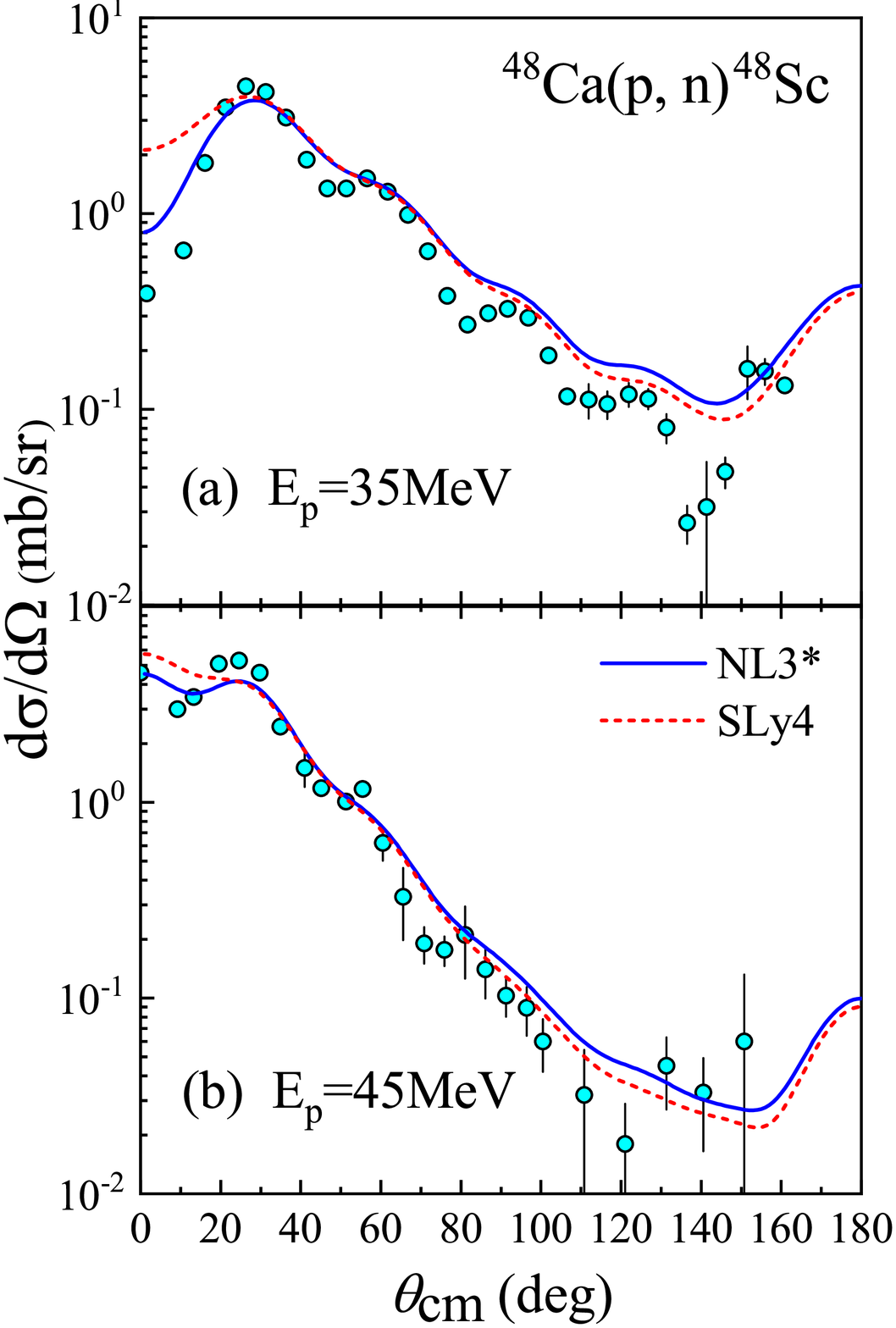}\\
	\caption{(Color online) Same as the Fig. \ref{fig:3}, but for the quasielastic $^{48}$Ca($p, n$)$^{48}$Sc reaction.}
	\label{fig:4}
\end{figure}

\begin{figure}[htbp]
	\centering
	\includegraphics[width=8.5cm,angle=0,clip=true]{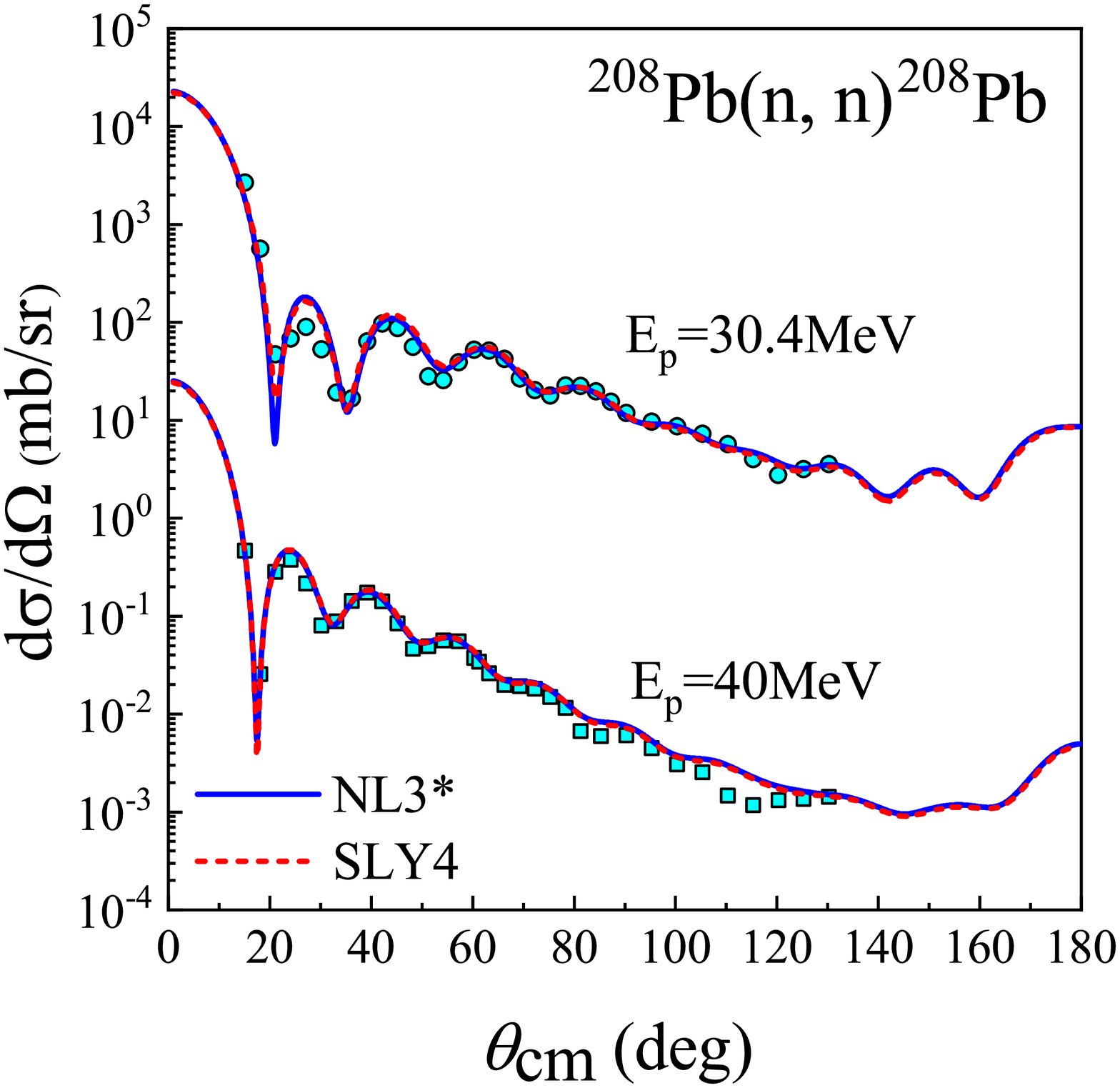}\\
	\caption{(Color online) Different ($n, n$) cross sections on $^{208}$Pb targets at 30.4$\,$MeV and 40$\,$MeV from the calculations with the complex folded potential of Eq. (\ref{26}), based on the nuclear densities calculated by the SHF and RMF models. The experimental data stem from Ref. \cite{devito2012neutron}.}
	\label{fig:nncomplex}
\end{figure}

Besides the $^{208}$Pb target, the theoretical cross sections for the $^{48}$Ca($p, n$)$^{48}$Sc reaction are also presented in Fig. \ref{fig:4}, using the renormalization coefficients in Table \ref{Table:III}. In the figure, one can again see that the general trend of theoretical results agrees with the experiment data, which supports the use of the renormalization coefficients. A further comparative study in Fig. \ref{fig:4} indicates that the NL3$^*$ results agree better with the ($p, n$) data than SLy4, especially in the forward direction. However, the renormalization coefficients are primarily based on the experimental result from PREX-II in the current paper. After the experimental $\Delta R_{n p}$ result of $^{48}$Ca is updated, more universal renormalization coefficients can be obtained, which are helpful for the analyses in this paper.

With the exception of the ($p, p$) and ($p, n$) scattering, the elastic neutron ($n, n$) scattering is also considered to prove the consistency of the folded potential. The elastic neutron ($n, n$) cross sections on $^{208}$Pb at 30.4$\,$MeV and 40$\,$MeV are shown in Fig. \ref{fig:nncomplex}. In analogy with the ($p, p$) scattering, the complex folded potential of Eq. (\ref{26}) is renormalized at different incident energies to obtain $N_{V} \approx 0.80$ and $N_{W} \approx 0.65\mbox{-}0.75$. From Fig. \ref{fig:nncomplex}, it can be seen that the complex folded potential of Eq. (\ref{26}) gives good ($n, n$) descriptions on cross section, which indicate the validity of the complex folded potential on ($n, n$) scattering. Therefore, our results demonstrate the consistency among the charge-exchange effective interaction, the proton and the neutron folded potential in our calculations.

The theoretical results in Figs. \ref{fig:3} and \ref{fig:4} together illustrate that the complex folded potential can reflect differences in $\rho_{n}$ on the ($p, n$) cross section. However, the renormalization coefficient $N_{V1}$ needs to be readjusted for different nuclei, which indicates that the complex folding model has some limitations as far as its universality is concerned.

\subsection{C. Hybrid folding model analysis}

The $N_{V1}$ factor of the complex folding model has been a function of the mass number $A$ of nucleus. To retreat in the renormalizations carried out from our side, we use the hybrid folded potential: 
\begin{equation}\label{29}
	\begin{aligned}
		U_{\mathrm{N}}(R)=N_{V}\left[V_{\mathrm{IS}}(R)-N_{V 1} V_{\mathrm{IV}}(R)\right]+i\left[W_{0}(R)-W_{1}(R)\right],
	\end{aligned}
\end{equation}
where the $V_{\mathrm{IS}}$ and $V_{\mathrm{IV}}$ terms retain the folded potential, and the imaginary part is replaced by that from a phenomenological optical model potential. Specifically in Eq. (\ref{29}), the $W_{0}(R)$ and $W_{1}(R)$ are the isoscalar and isovector parts of the imaginary Koning-Delaroche (KD) potential \cite{koning2003local}, respectively. The KD global systematics covers a wide range of target masses and energies. Similar to the case of the complex folded potential, we calibrate $N_{V(V1)}$ of the hybrid folded potential on the experimental ($p, p$) and ($p, n$) cross sections on $^{208}$Pb, assuming validity of the central value of the $\Delta R_{n p}$ PREX-II result, i.e., NL3$^{*}$ densities. The calibrated $N_{V(V1)}$ at the incident energies of 35$\,$MeV and 45$\,$MeV are given in Table \ref{Table:IV}. In this way, the renormalization coefficients are universal for different nuclei, but depend on energy.

\begin{table}[htbp]
	\begin{center}
		\caption{Renormalization coefficients $N_{V}$ and $N_{V1}$ of the hybrid folded potential of Eq. (\ref{29}) at 35$\,$MeV and 45$\,$MeV, which are calibrated based on the experimental ($p, p$) and ($p, n$) cross sections, assuming the validity of the central value of $\Delta R_{n p}$ from PREX-II.}
		\label{Table:IV}
\begin{tabular}{ccllllllllccclllllllllc}
	\hline\hline
	&  &  & $E$  &  &  &  &  &  &  &  &  & $N_{V}$    &  &  &  &  &  &  &  &  & $N_{V1}$   &    \\ \hline
	&  &  & 35 &  &  &  &  &  &  &  &  & 0.902 &  &  &  &  &  &  &  &  & 0.908 &    \\
	&  &  & 45 &  &  &  &  &  &  &  &  & 0.936 &  &  &  &  &  &  &  &  & 1.105 &    \\ \hline\hline
\end{tabular}
	\end{center}
\end{table}

\begin{figure}[htbp]
	\centering
	\includegraphics[width=8.5cm,angle=0,clip=true]{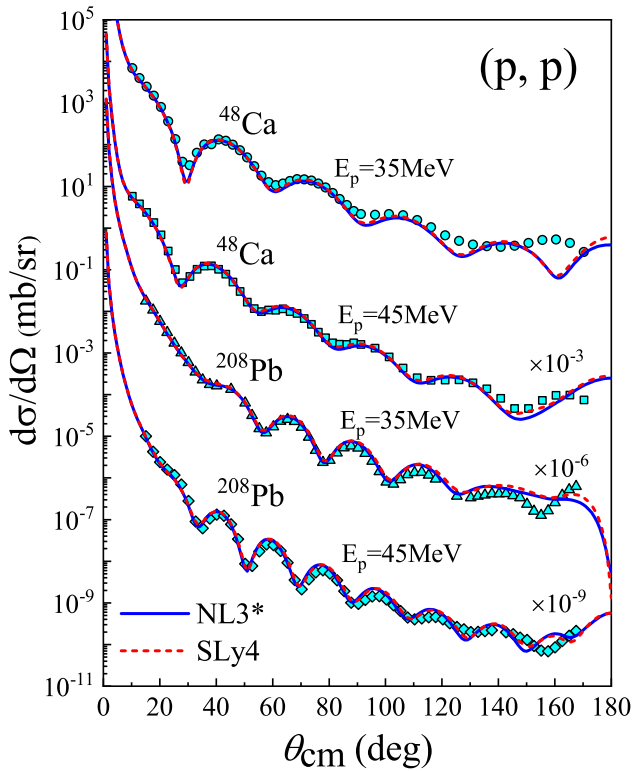}\\
	\caption{(Color online) Same as the Fig. \ref{fig:2}, but with calculations in the hybrid folded potential Eq. (\ref{29}).}
	\label{fig:5}
\end{figure}

The ($p, p$) cross sections on $^{48}$Ca and $^{208}$Pb, calculated with the hybrid folded potential Eq. (\ref{29}) and the renormalization coefficients in Table \ref{Table:IV} using NL3$^*$ and SLy4 interactions, are presented in Fig. \ref{fig:5}. The theoretical cross sections are in good agreement with the experimental data, which validates the use of the hybrid folding model with the calibrated renormalization coefficients. From Fig. \ref{fig:5} one can see that the effects of different $\rho_{n}$ on ($p, p$) cross sections are rather minute. This can be attributed to the fact that the impact of $V_{\mathrm{IV}}$ on the ($p, p$) cross sections is relatively small.
In comparing the results in Fig. \ref{fig:2} and Fig. \ref{fig:5}, one can observe clear differences between the ($p, p$) cross sections calculated with the complex folded potential and hybrid folded potential, especially in the backward region. These are due to the surface term of the imaginary part of isoscalar potential $U_{\mathrm{IS}}$. Notably, while the real part of the hybrid folded potential is quite close in shape and strength to the real KD potential, the imaginary part is quite different.  

\begin{figure}[htbp]
	\centering
	\includegraphics[width=8.5cm,angle=0,clip=true]{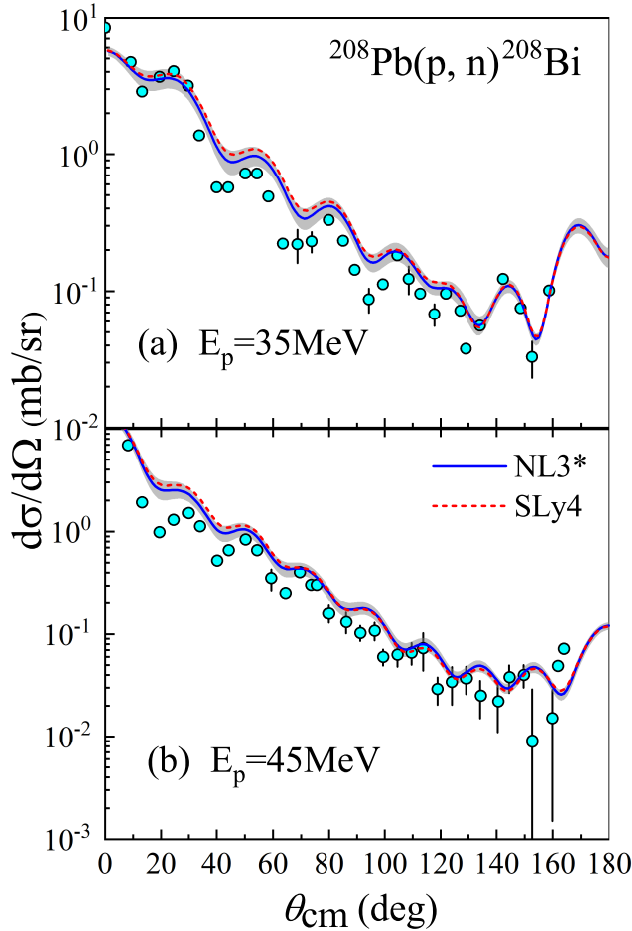}\\
	\caption{(Color online) Same as Fig. \ref{fig:3}, but for the hybrid folded potential of Eq. (\ref{29}).}
	\label{fig:6}
\end{figure}

With the renormalization coefficients of Table \ref{Table:IV}, the theoretical ($p, n$) scattering cross sections on $^{208}$Pb have been again calculated and are presented in Fig. \ref{fig:6}. By stretching neutron density $\rho_{n}$, the uncertainty in $R_{n}$ in the PREX-II measurement is again mapped onto the shaded areas. In comparing Fig. \ref{fig:3} and Fig. \ref{fig:6}, one can see that the hybrid folded potential gives better descriptions of the ($p, n$) data than the complex folded potential, which can be attributed to the surface term of the hybrid folded potential. Specifically, the imaginary KD potential can be represented by a combination of volume and surface terms. The imaginary folded potential only exhibits the volume character, since it is constructed based on the nucleon optical potential calculated by the nuclear matter. Therefore, the imaginary folded potential cannot appropriately explain the surface absorption of the transfer reactions caused by inelastic scattering and reflects only the nature of the volume \cite{khoa2007folding}. However, all phenomenological potentials have a surface-peaked form at low energies, which slowly changes to a volume form as the energy increases. In the range of incident energies studied in this paper, the surface absorption is still very strong. Thus, the ($p, n$) cross section given by the hybrid folded potential of Eq. (\ref{29}) is more accurate for the probe of the neutron density distribution.

In comparing the theoretical results from the NL3$^*$ and SLy4 interactions in Fig. \ref{fig:6}, it may be seen that the ($p, n$) cross sections predicted by the hybrid folded potential are also sensitive to $\rho_{n}$. This is because the transition strength of the ($p, n$) reaction to IAS is determined entirely by the isovector part in hybrid folded potential, although only the real part of isovector potential is now calculated from the derived nuclear density distribution. Therefore, even the hybrid folding model can also be used to study neutron density distribution $\rho_{n}$. Besides, the hybrid folding model may be viewed as a more objective inference method, since the renormalization coefficients are the same for different target nucleus. In the following, we progress to using the hybrid folding model in testing the impacts of the neutron properties of $^{48}$Ca.

\begin{figure}[htbp]
	\centering
	\includegraphics[width=8.5cm,angle=0,clip=true]{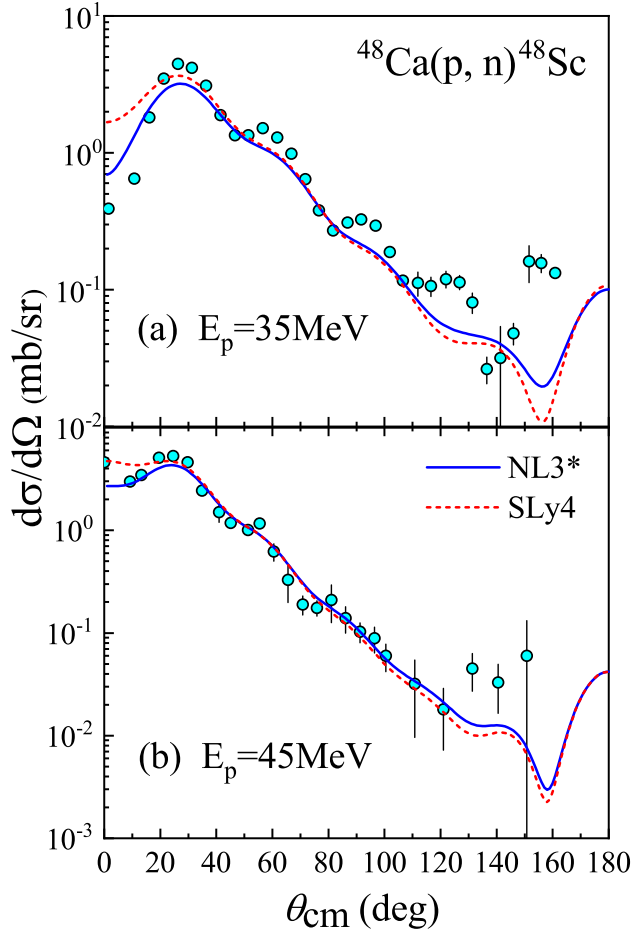}\\
	\caption{(Color online) Same as Fig. \ref{fig:4}, but for the hybrid folded potential of Eq. (\ref{29}).}
	\label{fig:7}
\end{figure}

In Fig. \ref{fig:7}, we present the ($p, n$) cross sections on $^{48}$Ca obtained in the hybrid folded potential at 35$\,$MeV and 45$\,$MeV, using the renormalization coefficients. It can be observed in this figure that the results from the hybrid folded potential provide good description of the $^{48}$Ca($p, n$)$^{48}$Sc quasielastic reaction data, which supports the universality of the renormalization coefficients in Table \ref{Table:IV}. In addition, we find that the ($p, n$) cross sections calculated with the NL3$^*$ and SLy4 interactions significantly differ in the regions $\theta =0^{\circ}\mbox{-}40^{\circ}$ and $\theta =80^{\circ}\mbox{-}160^{\circ}$. Therefore, we can effectively constrain the neutron properties following the hybrid folding model. In Fig. \ref{fig:7}, one can see that the results of NL3$^*$ parameter set are generally closer to the ($p, n$) data, especially in the forward and backward angles. This finding is consistent with the conclusions of the complex folding model analysis.

\section{\uppercase\expandafter{\romannumeral4}. Conclusion}

The neutron skin thickness $\Delta R_{n p}$ and the neutron density distribution $\rho_{n}$ are fundamental nuclear properties, which attracted increased attention recently. Relying on the relation between $\rho_{n}$ and the quasielastic ($p, n$) cross section in this paper, we have investigated the impact of neutron properties in the context of the available experimental values. 

In calculating the neutron properties in the RMF and SHF models, we found that the $\Delta R_{n p}$ and $\rho_{n}$ can differ significantly among different parameter sets. The elastic ($p, p$) and quasielastic ($p, n$) cross sections of $^{208}$Pb have been investigated in the combination of the DWBA method and the folding model. The renormalization coefficients for the folded potential have been calibrated using the experimental ($p, p$) and ($p, n$) data assuming central value of the neutron skin thickness $\Delta R_{n p}$ of $^{208}$Pb in the PREX-II measurement. The isovector potential determines the transition strength of the initial state to IAS in charge-exchange ($p, n$) reactions. Therefore, the accurate measurement of the ($p, n$) cross sections can serve as a sensitive probe of the neutron skin thickness $\Delta R_{n p}$ and the nuclear isovector density. Results in this paper also indicate that the ($p, n$) cross section is sensitive to the nuclear neutron density distribution $\rho_{n}$. By further comparing the results of the complex and hybrid folding model, we found that the ($p, n$) reaction can be more reasonable described by introducing the surface term into the folded potential.

With the renormalization coefficients calibrated in this paper, the ($p, n$) cross sections of $^{48}$Ca have also been calculated in the folding model for different neutron density distribution $\rho_{n}$. Theoretical quasielastic ($p, n$) cross sections have been compared with the experimental data. It has been observed that the results of NL3* parameter set are consistent with the experimental data. The results of this paper can provide counter reference for the CREX experiment. Besides, our investigations on charge exchange reactions are also helpful for other fields of nuclear structure and nuclear reactions.

\section*{Acknowledgements}

The authors are grateful to Dao T. Khoa for the valuable discussions and suggestions. This work was supported by the National Natural Science Foundation of China (Grants No. 11505292, No. 11822503, No. 11975167, and No. 12035011), by the Shandong Provincial Natural Science Foundation, China (Grant No. ZR2020MA096), by the Open Project of Guangxi Key Laboratory of Nuclear Physics and Nuclear Technology (Grant No. NLK2021-03), by the Key Laboratory of High Precision Nuclear Spectroscopy, Institute of Modern Physics, Chinese Academy of Sciences (Grant
No. IMPKFKT2021001), and by the U.S. Department of Energy, Office of Science under Grant No. DE-SC0019209.


\bibliography{references}
\end{document}